\documentstyle[12pt]{article}
\ifx\epsffile\undefined
\newlength{\epsfysize}
\def\epsffile#1#2#3#4]#5{}
\def\pageb{}
\def\clearp{}
\else
\def\pageb{\pagebreak}
\def\clearp{\clearpage}
\fi

\def\roughly#1{\raise.3ex\hbox{$#1$\kern-.75em\lower1ex\hbox{$\sim$}}}
\begin{document}
\begin{titlepage}
\begin{center}
February 1994              \hfill JHU-TIPAC-940001\\
			   \hfill PURD-TH-94-04\\
			   \hfill hep-ph/9403240\\
\vskip .7 in
{\large \bf The complete radiative corrections to the gaugino and \\
            Higgsino masses in the Minimal Supersymmetric Model}
\vskip .3 in
           \vskip 0.5 cm
      {\bf Damien Pierce}\\
      {\it Department of Physics and Astronomy\\
          The Johns Hopkins University\\
         Baltimore, Maryland\ \ 21218\\}
          \vskip 0.5 cm
      {\bf Aris Papadopoulos}\\
      {\it Physics Department\\
           Purdue University\\
           West Lafayette, Indiana\ \ 47907\\}
\end{center}
\vskip 0.4 in
\begin{abstract}
We determine the radiative corrections to the masses of the gauginos and
Higgsinos in the MSSM, including all sectors of the theory in a one-loop
calculation in the
on-mass-shell renormalization scheme. We find that a gluino which is
massless at tree level receives a mass of between 0 and 3~GeV, primarily
due to the top/stop contribution. This radiatively generated mass
depends directly on the off-diagonal element of the squark mass matrix.
In the case of a massive gluino, its mass
receives typically large corrections, as large as
40\% for a 125~GeV gluino. We find that the contributions to the
neutralino and chargino mass corrections from the
gauge/Higgs/gaugino/Higgsino sector are typically $\pm1\%$.
The lightest neutralino, which can receive corrections larger
than 25\%, receives
${\cal O}(5\%)$ corrections over most of the parameter space.
We combine our results with the results of LEP and CDF searches
to obtain the lower bounds on the neutralino and chargino masses
at one-loop.  We also demonstrate how the radiative corrections affect
the presently excluded region of parameter space.
\end{abstract}
\end{titlepage}
\renewcommand{\thepage}{\arabic{page}}
\setcounter{page}{1}

\section{Introduction}

In this paper we study the effects of radiative corrections on the
gaugino and Higgsino masses in the minimal supersymmetric model.
In a grand unified model the gaugino masses at the GUT scale are given
by the common mass $m_{1/2}$. Using the one-loop renormalization group
equations\cite{Falck}, the gaugino mass parameters $M_1,\ M_2,$ and
the gluino mass $M_3$ are given at a scale $\mu$ by
\begin{equation}
M_i(\mu^2) = {\alpha_i(\mu^2)\over\alpha_{\rm GUT}}m_{1/2}
\label{GUTrelation}
\end{equation}
which yields the relations $M_3\simeq3.5 M_2,$
and $M_1\simeq0.5 M_2$. We assume these GUT relations in this paper.
It is typical in analyses to assume that elements of the
gaugino/Higgsino sector are lighter than the lightest elements of the
squark/slepton sector, and
this is required if the dark matter candidate is the lightest
neutralino. In such a scenario, we might expect that members of the
gaugino/Higgsino sector would
be the first superpartners to be discovered at a particle collider.
It is therefore of interest to determine the corrections that
these particle masses receive. Until these particles are discovered,
the corrections are useful in that they modify the region of parameter
space which is ruled out by experiment. In the event that the gaugino
and Higgsino particles are discovered these corrections would provide for
a detailed check of the viability of the MSSM.

We divide the following discussion into four parts. In the next section we
discuss the radiatively generated mass for a gluino which is massless at
tree level. In Section 3 we discuss the corrections to the gluino mass
in the case $m_{\tilde{g}}\ \roughly{>}\ 100$~GeV. In section 4 we
discuss the complete corrections to the neutralino and chargino masses,
and the last section is reserved for the conclusions.

\section{Light gluino mass}

There is some controversy as to whether light gluinos
($m_{\tilde{g}}\ \roughly{<}$ 3~GeV) have been ruled out by existing
data\cite{PDG}. The bounds from the non-observation of the decay mode
$\Upsilon \rightarrow \gamma \eta_{\tilde{g}}$ \cite{CUSB}
are model dependent\cite{Yuan,Farrar} ($\eta_{\tilde{g}}$ is the
$\tilde{g}\tilde{g}$ pseudoscalar bound state). Ref.\cite{Farrar} concludes
that these searches exclude 3~GeV $\roughly{<}~m_{\eta_{\tilde{g}}}~
\roughly{<}~7$~GeV,
leaving the window $m_{\tilde{g}}\ \roughly{<}\ 1.5$~GeV open if
$m_{\tilde{g}}=m_{\eta_{\tilde{g}}}/2$. Bounds determined from beam dump
experiments\cite{BeamDump} depend on the gluino lifetime and may not rule
out very light gluinos if the squark masses are in the TeV range.
Thus, it appears further experiments are needed to conclusively rule out
the possibility of light gluinos, especially long lived gluinos with
mass $m_{\tilde{g}}<1$~GeV (see e.g. Ref.\cite{Sher}).
Interest in the possibility of light gluinos has been piqued recently
in response to the observation that a light gluino could reconcile the
apparent discrepancy between (central) values of $\alpha_s(M_Z^2)$ as
determined from high and low energy experiments\cite{Yuan,alphas}.

In this section we
examine the gluino mass generated entirely by one-loop radiative
corrections in the event $M_3$=0. At one-loop, the solution to the
RGE for the parameter
$M_3,\ M_3(\mu^2)=(\alpha_s(\mu^2)/\alpha_{\rm GUT})m_{1/2}$,
together with the GUT scale boundary condition $m_{1/2}$=0
would ensure $M_3$=0. Considering the two-loop RGE for the gluino
mass\cite{two-loopRGE},
we see that if both $m_{1/2}$ and the $A$-terms vanish at the GUT
scale, again we would find a massless gluino at tree level.
In any event, if the parameter $M_3(M_3^2)$ is small
({$\roughly{<}$}\ few GeV) then the correction given below can
simply be added to it. The contribution to the gluino mass for
$M_3=0$ due to the top/stop loop is
\begin{equation}
m_{\tilde{g}} = {\alpha_s(m_t^2)\over4\pi}m_t\sin2\theta_t\left(
{\ln r_1\over 1-r_1} - {\ln r_2 \over 1-r_2} \right),\end{equation}
where $r_{1,2}=m_t^2/m_{\tilde{t}_{1,2}}^2$ and $\theta_t$ is the
angle which rotates the squarks from the
left-right to the mass eigenstate basis.
Changing the scale at which the strong coupling is evaluated from
$m_t$ (160~GeV) to $m_{\tilde{t}}$ (1 TeV) reduces the gluino mass
by 14\% (we set $\alpha_s(M_Z^2)=0.12$). We sum over the six quarks,
but only the
top/stop contribution matters except at large values of $\tan\beta$,
where the bottom/sbottom contribution can be non-negligible.
We note here that our result for the gluino mass, as well as the photino mass,
is larger by a factor of 2 relative to those of Refs.
\cite{Yuan,Barbieri}. We believe that our result is correct since
the cancellation of infinities in the gluino and neutralino masses
depends on it.

For a given flavor, the squark masses and rotating angle $\theta$
are determined
from the squark mass matrix. For up type squarks, for example, we have
\begin{equation}
\bf M^2_{\rm sq}=
\left( \begin{array}{cc} a_u & c_u\\ c_u & b_u \end{array} \right)
\end{equation} where \begin{eqnarray}
a_u &=& M_{\tilde{Q}}^2+M^2_Z\cos 2\beta \left(1/2-2/3 s^2_w\right)+m^2_u\\
c_u &=& m_u(A_u-\mu\cot \beta)\\
b_u &=& M_{\tilde{U}}^2+ 2/3 M^2_Z\cos 2\beta s^2_w  +m^2_u
\end{eqnarray}
We prefer to present our results in terms of the underlying parameters
$M_{\tilde{Q}},\ M_{\tilde{U}},\ \mu,$ and $A$. We discuss the results
for a common $A$-term, $A=A_t=A_b$, and a common right-handed squark mass
$M_{\tilde{U}}=M_{\tilde{D}}$. We set the top quark mass to 160~GeV.
In Fig.1(a)  we show gluino mass contours in the
$M_{\tilde{Q}},\ \mu$ plane, with $\tan\beta\!=\!1.5,$ $M_{\tilde{U}}\,$
=$\,M_{\tilde{D}}\,$=\,200~GeV, and $A$=0. We consider
the range $|\mu|<800$~GeV, and $100 < M_{\tilde{Q}} < 1000$~GeV.
We note that the gluino mass is at most about 3~GeV. It is equal to
0 near $\mu=A\tan\beta$, since in this case there is no mixing
between left and right stop squarks. The mass increases as we move away from
the no-mixing line until we reach the boundary of the parameter space,
defined by the requirement that the squared squark masses are positive.
The $\times$'s on Figs.1(a,c,d) mark unphysical regions of parameter space.
As we increase $M_{\tilde{U}},M_{\tilde{D}}$
the allowed parameter space increases and
\begin{figure}[htb]
\epsfysize=4.33in
\epsffile[65 400 610 730]{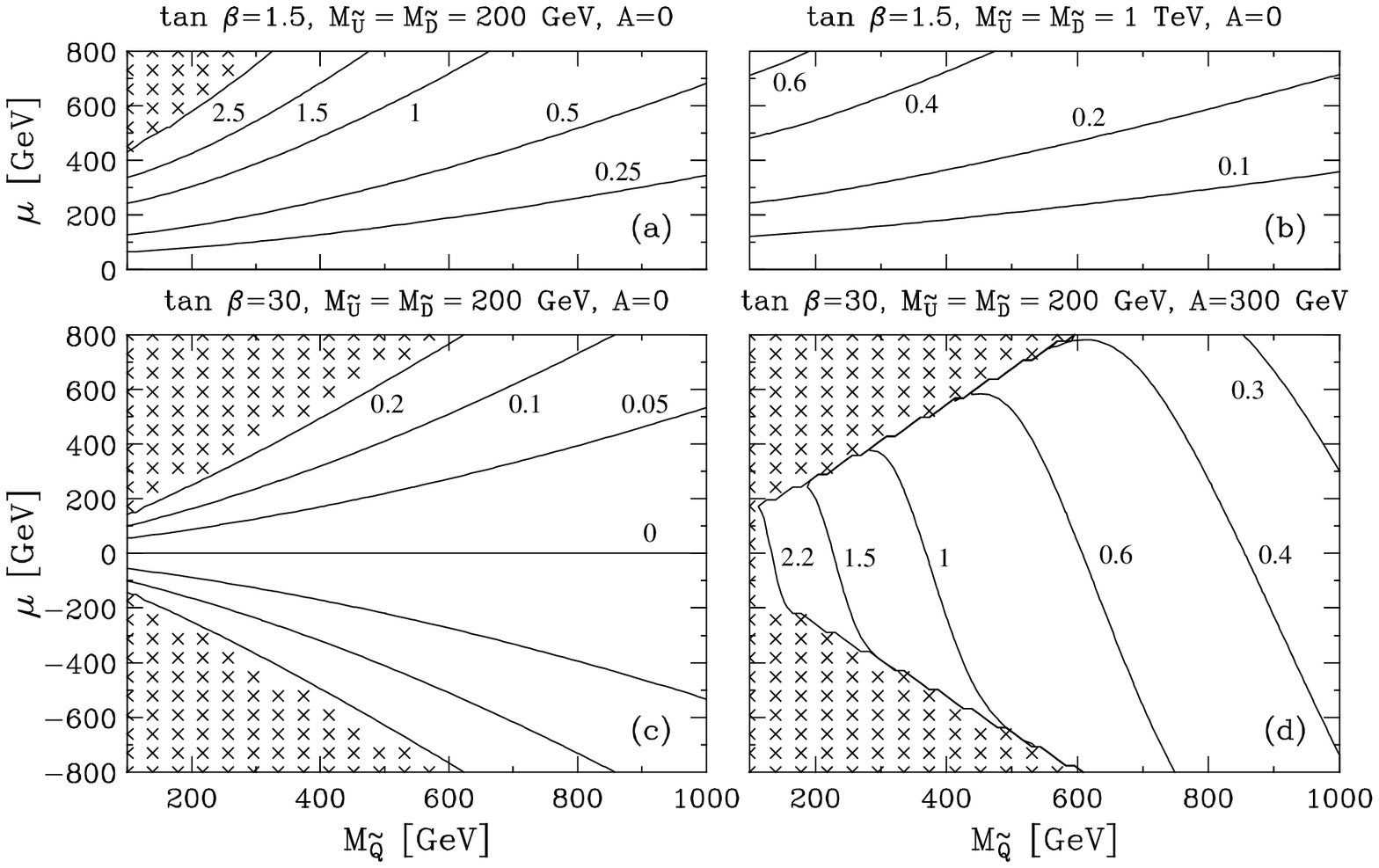}
\begin{center}
\parbox{5.5in}{
\caption[]{\small Contours of the radiatively generated gluino mass
for a gluino which is massless at tree level,
in the $M_{\tilde{Q}},\ \mu$ plane.
In Figs.(a,b,c) the contours are symmetric in
$\mu\rightarrow-\mu$. The contours are labeled in GeV.
The $\times$'s mark unphysical regions of parameter space.}}
\end{center}
\end{figure}
the bound on the gluino mass generally decreases. For example,
in Fig.1(b) $M_{\tilde{U}}$ and $M_{\tilde{D}}$ are increased to 1~TeV,
and the maximum gluino mass is about 700 MeV.
The dependence of the contour plots in Figs.1(a,b) on $\tan\beta$ and
$A$ is easily understood, since for small $\tan\beta$
the gluino mass depends directly on the parameters
$A,\ \mu,$ and $\tan\beta$ through the combination
$A-\mu\cot\beta$. Therefore changing $A$ or $\tan\beta$ amounts to simply a
shift or rescaling of the y-axis of these figures. For
large $\tan\beta$ the dependence
is more complicated, since the bottom-sbottom sector becomes
non-negligible,
and the parameter space is more severely constrained. In Fig.1(c) we
show the gluino mass contours for
$\tan\beta=30,\ M_{\tilde{U}}\,$=$\,M_{\tilde{D}}\,$=200~GeV,
and $A=0$. The gluino mass in this case varies from 0 at $\mu$=0 to
about 400~MeV. Again, the bound is reduced if we increase $M_U,
M_D$. For $M_U$=$M_D$=1 TeV and $\tan\beta=30$ the gluino mass
is at most 120 MeV.
As we vary $A$ when $\tan\beta$ is large the pattern of the gluino mass
contours in the $M_{\tilde{Q}},\ \mu$ plane changes dramatically, as can
be seen in Fig.1(d), where $A$ is increased to 300~GeV.
Finally, we note that if we instead examine
contours in the $M_{\tilde{U}},\ \mu$, plane keeping
$M_{\tilde{Q}}$ constant, we get essentially the same results.  This is
due to the fact that in $a_u, b_u$ defined above,
$M_{\tilde{Q}}$ and $M_{\tilde{U}}$ dominate the contribution of the
D-terms  which are proportional to $M^2_Z$. This is true especially
at $\tan\beta\simeq 1$ but it persist for larger values of $\tan\beta$ as well.
\section{Heavy gluino mass}
In this section we consider radiative corrections to the gluino mass
in the case $M_3\neq0$. Naturally the gluino mass limits referred to in
this section assume that light gluinos are excluded.
The CDF collaboration obtained the bound $m_{\tilde{g}}>141$~GeV
under the unrealistic assumption that the branching ratio
$B(\tilde{g}\rightarrow\chi_1^0 q\bar{q})=1$ \cite{CDF}.
They also included the effects of cascade decays\cite{Baer_cascade}
in their analysis and, for a particular choice of
parameters, obtained a bound $m_{\tilde{g}}>95$~GeV \cite{CDF}. If the
unification condition for the gaugino masses is assumed, combining the
LEP limits on chargino and neutralino masses\cite{LEP} with the gluino
cascade analysis further constrains the gluino mass bound (and the bound
then becomes $\mu$ and $\tan\beta$ dependent). Such an analysis
was carried out in Ref.\cite{Baer}, where, for a typical choice of
parameters, they obtained $m_{\tilde{g}}>135$. Hidaka \cite{Hidaka}
performed a similar analysis, obtaining the $\mu$ and $\tan\beta$
independent bound $m_{\tilde{g}}>132$. However, the starting point for
both of these last two reference's analysis was $m_{\tilde{g}}>150$~GeV
with $B(g\rightarrow\chi_1^0 q\bar{q})=1$, a preliminary CDF
result. Hence, these bounds should be relaxed
accordingly. A reasonable lower bound on the gluino mass is
$m_{\tilde{g}}>125$~GeV \cite{Haber}.

The one-loop corrected gluino mass is given by
$$ m_{\tilde{g}} = M_{3}(\mu^2) + {3\alpha_s\over
4\pi}M_3\left(5-3\ln\left({{M_3^2\over\mu^2}}\right)\right)$$
\begin{equation}
-\sum_{q=u,..,t}
{\alpha_s\over4\pi}M_3\,{\rm Re}\left[
 \hat{B}_1(M_3^2,m_q^2,m^2_{\tilde{q}_1})
+\hat{B}_1(M_3^2,m_q^2,m^2_{\tilde{q}_2})\right]
\label{gluino mass}\end{equation}
$$ +\sum_{q=t,b}{\alpha_s\over4\pi}m_q\sin2\theta_q\,{\rm Re}\left[
B_0(M_3^2,m_q^2,m^2_{\tilde{q}_1})-
B_0(M_3^2,m_q^2,m^2_{\tilde{q}_2})
\right], $$
where we evaluate $\alpha_s$ at the scale $M_3$.
We take the tree level gluino mass to be $M_3(M_3^2)$.
The functions $B_0$ and $B_1$ can be found in Ref.\cite{Passarino},
except that here we adopt the metric $(+,-,-,-)$. The hatted $B_1$'s
refer to the function $B_1$ with the infinite part
(proportional to $1/\epsilon + \ln4\pi - \gamma_E$) subtracted.
We list in Eq.(\ref{gluino mass})
the gluon contribution, followed by the quark/squark
contributions. In the first sum over quarks we
include all six flavors, while in the second we indicate that the sum
is relevant only for $t$ and $b$ quarks. We note that the one-loop level
$\mu$ dependence of the parameter $M_3(\mu^2)$ in Eq.(\ref{gluino mass})
cancels against the $\mu$ dependence of the one-loop correction.

Given a measurement of the physical gluino mass,  we are interested in
determining the value of the underlying parameter $M_3$, which will then
yield a prediction for the other gaugino mass parameters $M_1$ and
$M_2$ via eq.(\ref{GUTrelation}).
To this end we plot in Fig.2(a) $M_3$ vs. $m_{\tilde{g}}$, for $m_t$=160
GeV. For the remainder of this paper, we will refer to a common squark mass
parameter $m_{sq}$, defined by
$m_{sq}=M_{\tilde{Q}}=M_{\tilde{U}}=M_{\tilde{D}}=2A$, where $A$ is the
common $A$-term.
\begin{figure}[htb]
\epsfysize=3in
\epsffile[65 360 610 600]{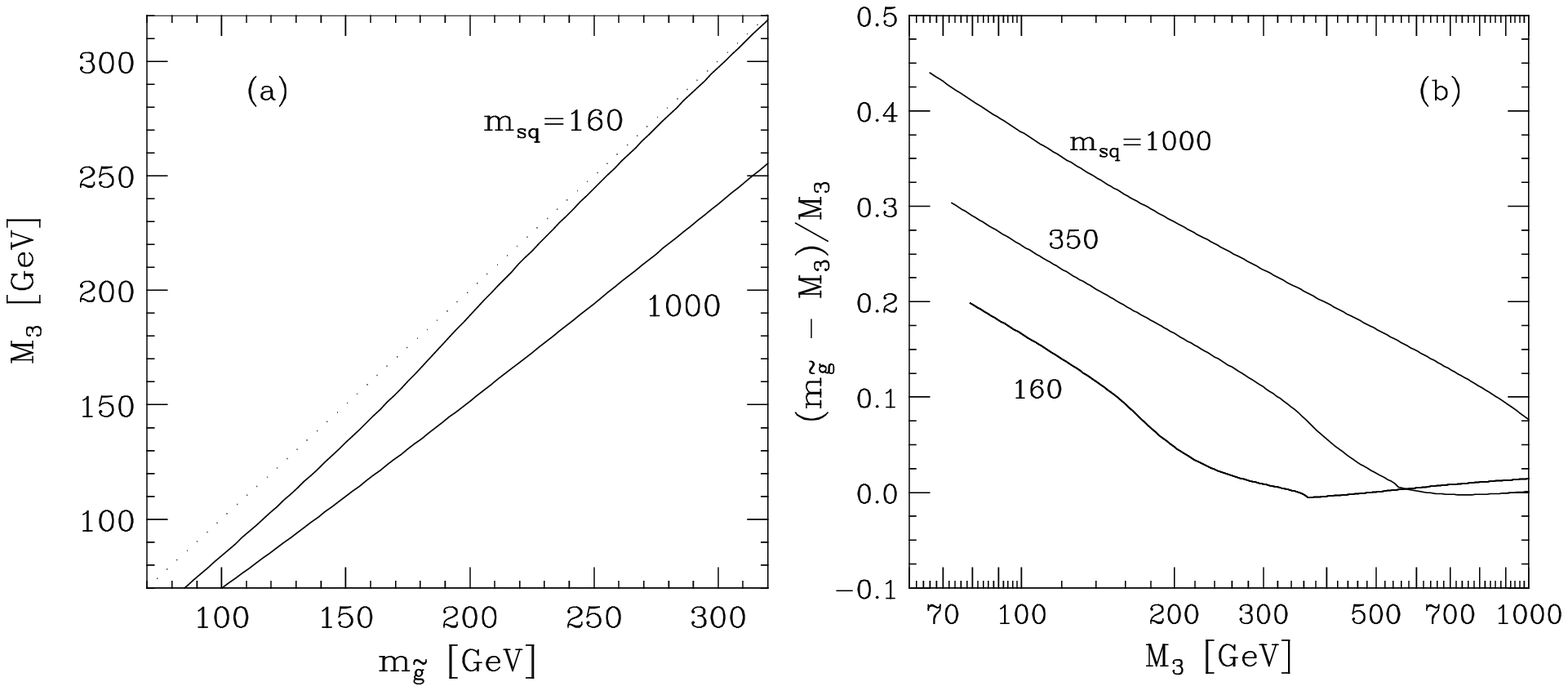}
\begin{center}
\parbox{5.5in}{
\caption[]{\small (a) The tree level gluino mass $M_3$ vs.
the one-loop gluino mass
$m_{\tilde{g}}$ for $m_t=160$~GeV and two values of the
squark mass, $m_{sq}= 160$~GeV and $m_{sq}=1000$~GeV. The dotted line
indicates the tree level relation $m_{\tilde{g}}=M_3$.
(b) The relative difference between $m_{\tilde{g}}$ and $M_3$, vs.
$M_3$ for three values of the squark mass.}}
\end{center}
\end{figure}
(In the plots of Figs.2 we set $\tan\beta$=1.5 and $\mu=-200$~GeV, but
the results are insensitive to the values of these parameters).
We see that a measured mass $m_{\tilde{g}}=100$~GeV would imply
$M_3$=69~GeV (84~GeV) for $m_{sq}=1000$~GeV (160~GeV). This is to be compared
with the tree level relation $m_{\tilde{g}}=M_3$.
This amounts to a 15\% to 40\%
correction to this relation for gluino masses at the present
limit of 125~GeV. In a grand unified context the gluino mass would
typically be taken to be the parameter $M_3$ evaluated at the scale
$M_3$. We show the error in neglecting the low energy
threshold corrections in this case in
Fig.2(b), where we plot the relative correction
to the gluino mass vs. $M_3$ for three values of the squark mass
parameter $m_{sq}$. The left endpoints of the curves in Fig.2(b)
correspond to $m_{\tilde{g}}=100$~GeV. For values of
$M_3$ less than the quark/squark thresholds the corrections are
generally appreciable. However, once $M_3$ is larger than the
quark/squark thresholds the relative correction to the tree level mass
is small ($\sim$1\%) and asymptotically approaches the value
$(3\alpha_s)/(4\pi)$.

\section{Chargino and neutralino masses}

In the MSSM the superpartners of the neutral Higgs bosons and
$SU(2)\times U(1)$ gauge bosons mix to form the mass eigenstates
(in order of increasing mass) $\chi_1^0,\ \chi_2^0,\ \chi_3^0,$
and $\chi_4^0$. Similarly, the charginos are the mass eigenstates of
the charged gauginos and Higgsinos, and are denoted $\chi_1^+$
and $\chi_2^+$ (see Ref.\cite{Haber&Kane}).
A lower bound on the lightest neutralino mass in
the MSSM was first obtained using combined LEP and UA2/CDF data in
Ref.\cite{Ros}, and more recently lower bounds on all of the neutralino
and chargino
masses were given in Ref.\cite{Hidaka}. The corrections to the
neutralino and chargino masses were studied in Ref.\cite{Lahanas}
including only the top/stop sector contributions, and in
Ref.\cite{ourpaper}, including the lepton/slepton and quark/squark
contributions. In this section we present the the complete corrections to
the chargino and neutralino masses, including the
gauge/Higgs/gaugino/Higgsino
sector, thus completing the calculation of Ref.\cite{ourpaper}.
In Ref.\cite{ourpaper} we found the corrections due to matter loops
were commonly of order 5\%. As the corrections due to
the gauge/Higgs/gaugino/Higgsino sector are expected to be of
this same order, they must be included in order to obtain a reliable result.

For consistency we use the same formalism as in Ref.\cite{ourpaper} which
we shall not repeat here. We point out one inconsequential modification,
which we adopt for technical reasons. Specifically, when computing the
masses, defined as the poles of the relevant two-point Greens
functions, we need to determine  counterterms for the various input
parameters that enter
the calculation. One of these counterterms is $\delta\beta$, where
$\tan\beta$ is the ratio of the the two vevs of the Higgs fields $H_1$
and $H_2$. In our renormalization scheme we take  $\delta\beta$ to be
purely infinite, i.e.
proportional to $1/\epsilon$ in dimensional regularization.
Since $\delta\beta$ is purely infinite, it will not contribute to the final
finite result. Yet, it is important to have an expression for it in order
to check the crucial cancellation of infinities, which must occur in
a renormalizable theory. In Ref.\cite{ourpaper} we determined
$\delta\beta$ from the Higgs sector. In the present  work we determine
it from the chargino-neutralino sector. In particular, we demand that
the the chargino masses, as well as the trace of the eigenvalues and the
squares of the eigenvalues of the neutralino mass matrix, are finite. This
determines the counterterms
$\delta M_1,\ \delta M_2,\ \delta\mu$ and $\delta\beta$. Notice that even
after imposing finiteness of these quantities it is still nontrivial
to check that all of the individual neutralino masses are finite. In
addition, we explicitly checked that we get the same value for
$\delta\beta$ whether we calculate it from the Higgs sector or the
chargino-neutralino sector.

The self-energies of the gauge
bosons, which enter in the computation of the counterterms, can be
found in Ref.\cite{Chank}\footnote{The definition of the gauge boson
self-energies given in Ref.\cite{ourpaper} differ by a sign relative to
those of Ref.\cite{Chank}. Similarly, our definition of the function $B_1$
differs by a sign from Ref.\cite{Chank}. We point out two
typographical errors in the formulas for the gauge boson self-energies
given in Ref.\cite{Chank}. First, in the formula for $\Pi_Z^T$, Eq.(A.15),
$s_W^2$ in the term $-2g_2^2s_W^2M_Z^2b_0(M_W^2,M_W^2)$ should be
replaced by $s_W^4$. Second, in the equation for $\Pi_W^T$, Eq.(A.20),
the $g_2^2$ in the term $-{1\over4}g_2^2\sum_{i=1}^2
(C_R^i)^2b_0(M_W^2,m_{H_i^0}^2)$ should be $g_2^4$.}.
The physical chargino and neutralino masses are given by \cite{Aoki}
\begin{equation}
{M_{\chi^+_i}} =  M_{\chi^+_{ir}} + \delta M_{\chi^+_i}
                -{\rm Re}\left(\Sigma^+_{1_{ii}}(M_{\chi^+_i}^2)+
              M_{\chi^+_i} \Sigma^+_{\gamma_{ii}}(M_{\chi^+_i}^2)\right)
\end{equation}
\begin{equation}
{M_{\chi^0_{\alpha}}} =
            M_{\chi^0_{\alpha r}} + \delta M_{\chi^0_{\alpha }}
      -{\rm Re}\left(\Sigma^0_{1_{\alpha \alpha}}(M_{\chi^0_{\alpha}}^2)+
M_{\chi^0_{\alpha}} \Sigma^0_{\gamma_{\alpha \alpha}}
 (M^2_{\chi^0_{\alpha}})\right)
\end{equation}
where the $\Sigma$'s are form factors of the one-loop fermion
self-energy
\begin{equation}
\Sigma_{ij}=
\Sigma_{1_{ij}}+\Sigma_{5_{ij}}\gamma_5
+\Sigma_{\gamma_{ij}}\rlap /p +\Sigma_{5\gamma_{ij}}\rlap /p\gamma_5,
\end{equation}
and the $M_{\chi_r} + \delta\!M_\chi$ are the masses obtained by
replacing each bare parameter $x_{i_b}$ in the tree level neutralino and
chargino mass formulas
by the renormalized parameter plus the corresponding counterterm,
$x_{i_b}\rightarrow x_{i_r} + \delta x_i$. (Here the $x_i$ refer to
the parameters $M_W$, $M_Z$, $M_1$, $M_2$, $\mu$, and $\tan\beta$; see
Ref.\cite{ourpaper} for details.) The gauge/Higgs/gaugino/Higgsino
contributions to the relevant chargino and neutralino self-energies
are given in the Appendix.

We set the top mass to 160~GeV and the {\it CP}-odd Higgs boson mass
$m_A$ to $m_{sq}/2$ in the following. The results depend mildly on the
scale of the Higgs boson sector, especially for any $m_A>M_Z$.
The tree level masses are functions of the underlying parameters
$M_1,\ M_2,\ \mu$ and $\tan\beta$. In our renormalization scheme these
parameters are identified with the $\overline{\it DR}$-running parameters
$M_1(Q^2),$ etc., and hence they have a renormalization scale dependence.
While the $Q$ dependence cancels in the result for the physical masses,
the corrections ($\Delta M_\chi$ or $\Delta M_\chi/M_\chi$)
do depend on the scales chosen for the evaluation of the tree level
masses. We
\begin{figure}[h]
\epsfysize=7.6in
\epsffile[30 60 400 730]{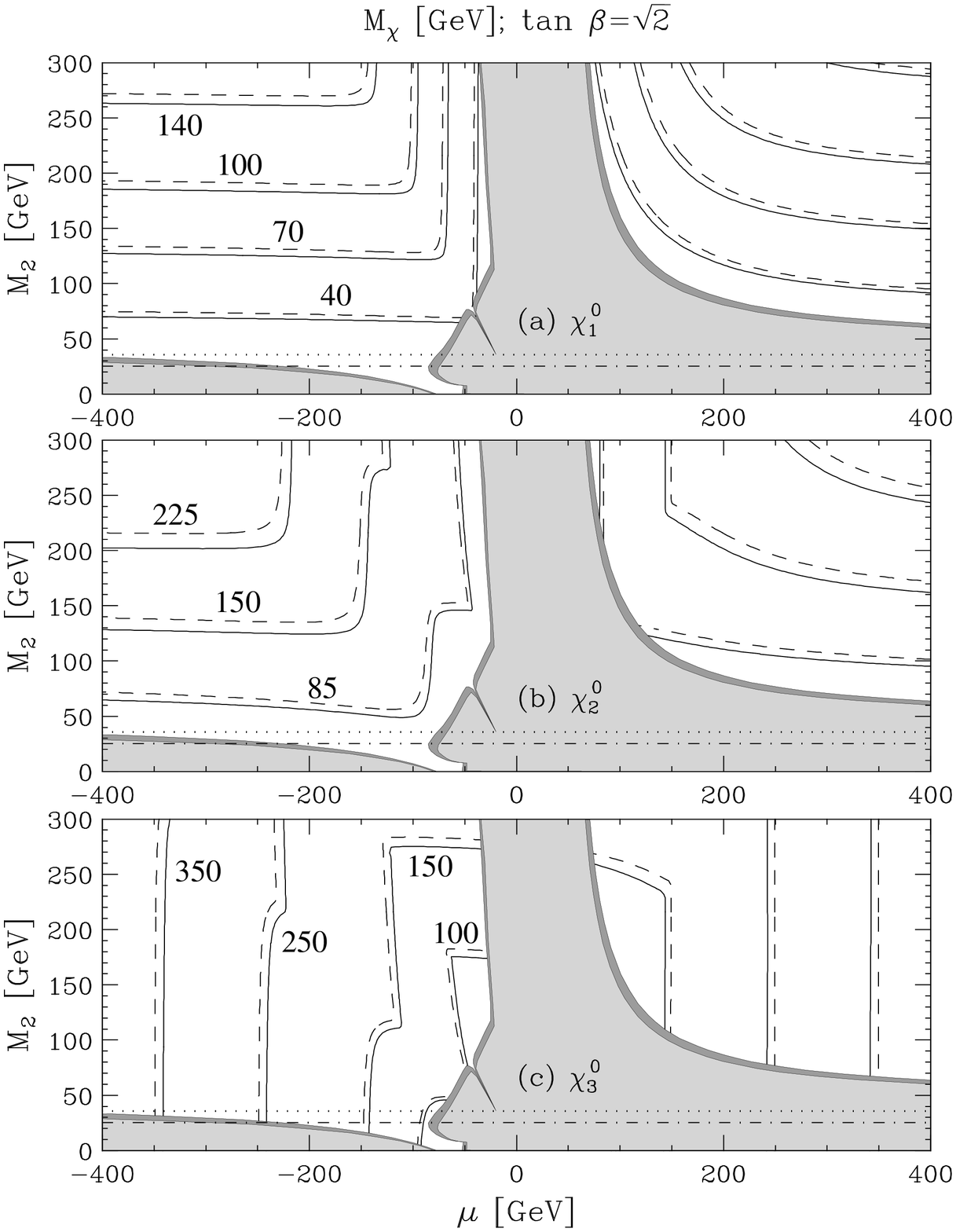}
\begin{center}
\parbox{6.3in}{
\caption[]{\small Contours of constant neutralino mass in the
$\mu,\ M_2$ plane with $\tan\beta=\sqrt{2}$ at tree level (dashed lines)
and one-loop level (solid lines). The light (dark + light)
shading indicates the region of parameter space presently ruled out by
LEP, including (not including) the one-loop corrections.
Additionally the region below the dotted (dash-dotted)
line is the tree (one-loop) level region ruled out by CDF gluino
searches. The contours are labeled in GeV.}}
\end{center}
\end{figure}
\begin{figure}[h]
\epsfysize=7.8in
\epsffile[55 60 400 730]{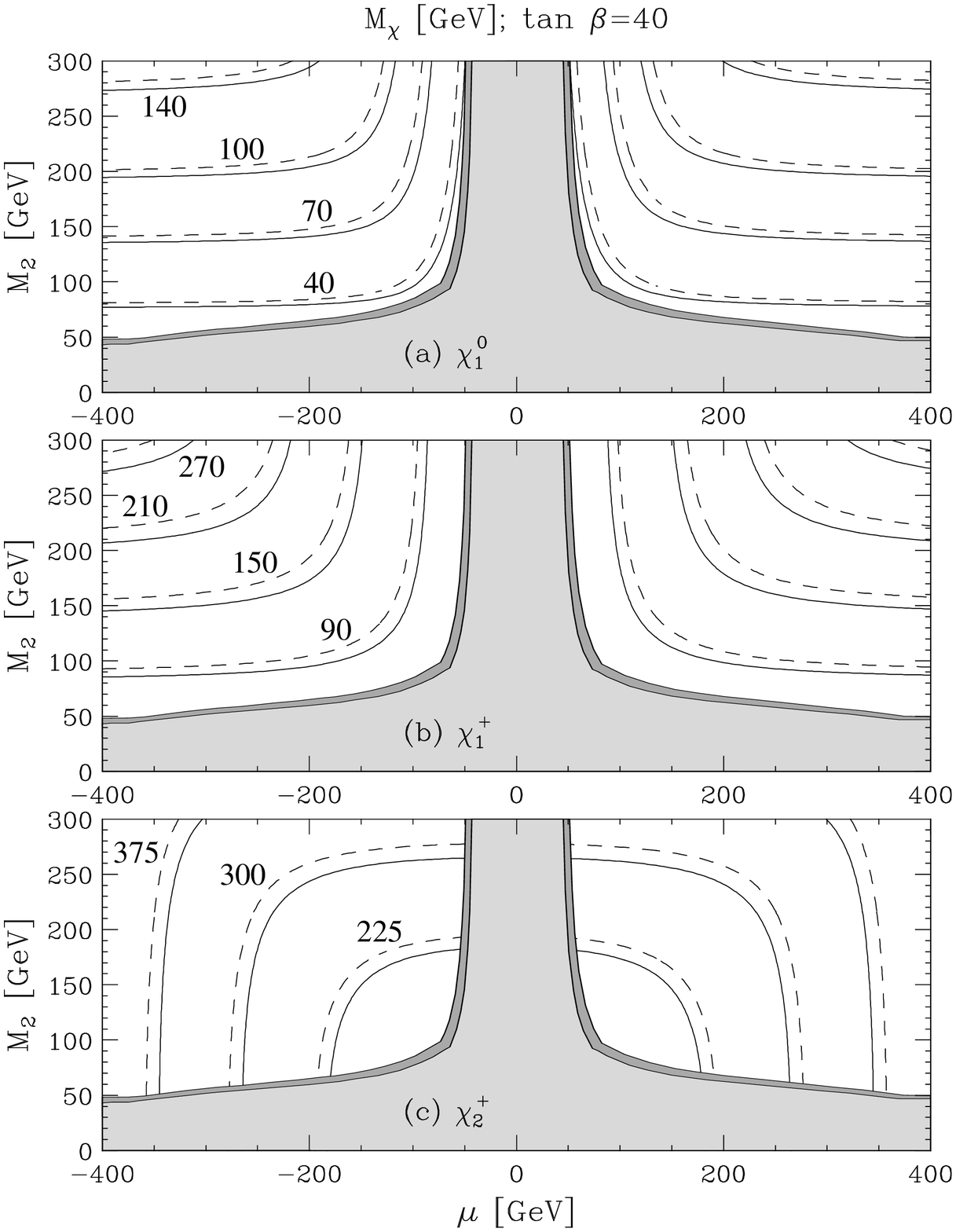}
\begin{center}
\parbox{5.5in}{
\caption[]{\small Same as Fig.3, with $\tan\beta=40$. The mass contours are
shown for (a) the lightest neutralino, (b) the lighter chargino,
and (c) the heavier chargino. In this case the gluino mass bound
gives no additional constraint.}}
\end{center}
\end{figure}
\clearp
\noindent choose to evaluate the tree level masses with the parameters
$M_1(M_1^2),\ M_2(M_2^2),\ \mu(\mu^2),$ and $\tan\beta(M_Z^2)$.

In Figs.3(a-c) we plot the  mass contours of
$\chi^0_1,\ \chi^0_2$ and $\chi^0_3$ in the $\mu,\ M_2$ plane at tree
(dashed lines) and one-loop (solid lines) level, for
$\tan\beta=\sqrt{2}$ and $m_{sq}=1$~TeV. In Figs.4(a-c) we
show the mass contours of $\chi^0_1,\ \chi^+_1,$ and $\chi^+_2$, for
$\tan\beta=40$. The shaded regions on these
figures indicate the
excluded regions of parameter space due to negative searches and
line-shape measurements at LEP\cite{LEP}.
Because the corrections are in general positive, there is a smaller
region of parameter space ruled out at one-loop level.
The light shaded regions shown in Figs.3 and 4 are ruled out at
one-loop level, while the adjacent strips
of dark shading indicate the additional region which is ruled out
if the chargino and neutralino masses are considered at tree level.
The boundary of the excluded region is determined largely from
the bound $M_{\chi^+_1}>47$~GeV \cite{LEP}.
The region below the dotted (dash-dotted) horizontal line in Figs.3
is excluded at tree level (one-loop level) by considering the
non-observation of gluinos by CDF and assuming the GUT relation
Eq.(\ref{GUTrelation}). The horizontal lines on the plots
correspond to $m_{\tilde{g}}>125$~GeV,
which translates by
Eqs.(\ref{GUTrelation},\ref{gluino mass}) into $M_2>36$~GeV (25~GeV)
at tree level (one-loop level). The bound $m_{\tilde{g}}>125$~GeV
provides no additional excluded region
beyond the region that is already excluded by LEP
for $\tan\beta\ \roughly{>}\ 2$.

We note that for a given $\tan\beta$
the results for the lightest chargino and neutralino are qualitatively
similar, and the same is true for the heaviest chargino and neutralino.
For $\tan\beta=\sqrt{2}$ and $\mu=-200$~GeV
a measured $\chi^0_1$ mass of 100~GeV would imply $M_2=192$~GeV if the tree
level relation is used, while taking into account radiative corrections
we obtain $M_2=185$~GeV. Generically, the corrections $\Delta M_\chi$
are below 10~GeV (12~GeV) in the region $M_2\ \roughly{<}\ 120$~GeV,
$|\mu|\ \roughly{<}\ 200$~GeV at $\tan\beta$=$\sqrt{2}$ ($\tan\beta$=40)
and they increase as we move away from the excluded region. For the
next-to-lightest neutralino and lightest chargino
with $|\mu|\simeq350$~GeV and
$M_2\simeq250$~GeV we can obtain a correction $\Delta M_\chi$
greater than 14~GeV.

\begin{figure}[htb]
\epsfysize=3.5in
\epsffile[65 478 680 758]{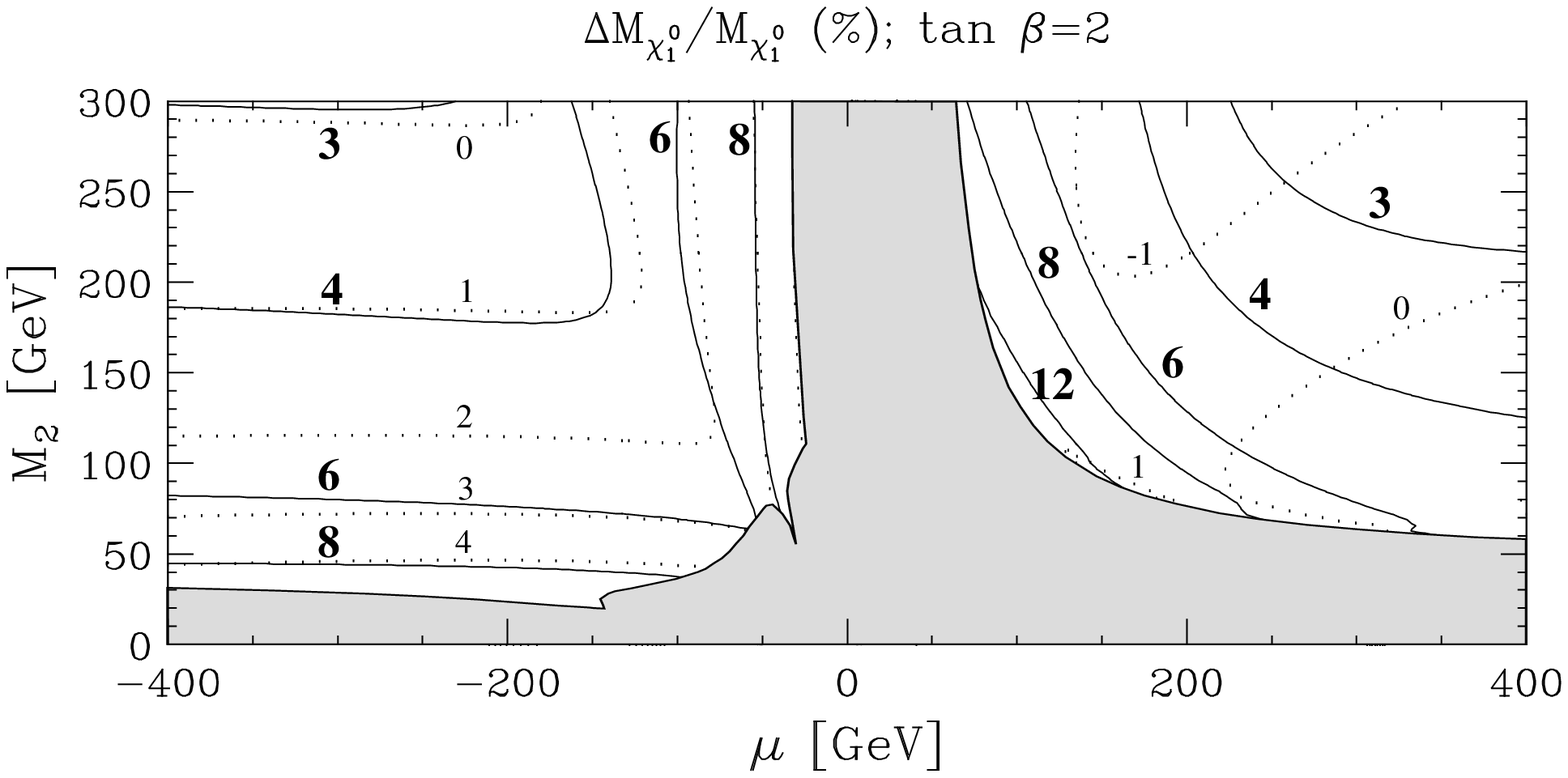}
\begin{center}
\parbox{5.5in}{
\caption[]{\small Contours of the percentage difference between the tree
and one-loop level values of the lightest neutralino mass in the
$\mu,\ M_2$ plane with $\tan\beta=2$. The solid lines (with large labels)
indicate the the results for $m_{sq}=1$ TeV, while the $m_{sq}=180$~GeV
results are indicated with dotted lines (small labels). The shading
indicates the one-loop level excluded region with $m_{sq}=1$ TeV.}}
\end{center}
\end{figure}
We illustrate that the corrections depend strongly on the squark
mass in Fig.5, where we show a contour plot of the percentage correction for
the lightest neutralino mass in the $\mu,\ M_2$ plane for $\tan\beta=2$
and $m_{sq}=1$~TeV (solid lines),
as well as for $m_{sq}=180$~GeV (dotted lines). Also shown is the
excluded region at one-loop with $m_{sq}=1$~TeV. We see that
in most of the parameter space the correction is
$3\sim8\%$ for $m_{sq}$=1~TeV, and $-1\sim4\%$ for
$m_{sq}=180$~GeV. For $m_{sq}=1$ TeV  the corrections are
significantly larger for smaller $\tan\beta$ or larger $m_t$.
The 12\% contour line shown in Fig.5 increases to 23\% if
$\tan\beta=1$.

Generally, the relative corrections are smaller for the heavier particles.
For example, with $\tan\beta=\sqrt{2}$ and $m_{sq}$=1~TeV we obtain,
in the region of parameter space shown in Fig.5, a correction
between 5\% and 15\% for $\chi^+_1$, a 3-7\% correction
for $\chi_2^0$, and for $\chi_3^0,\ \chi_4^0$
and $\chi_2^+$ a correction of 3-4\%.

\begin{figure}[htb]
\epsfysize=3in
\epsffile[65 360 610 600]{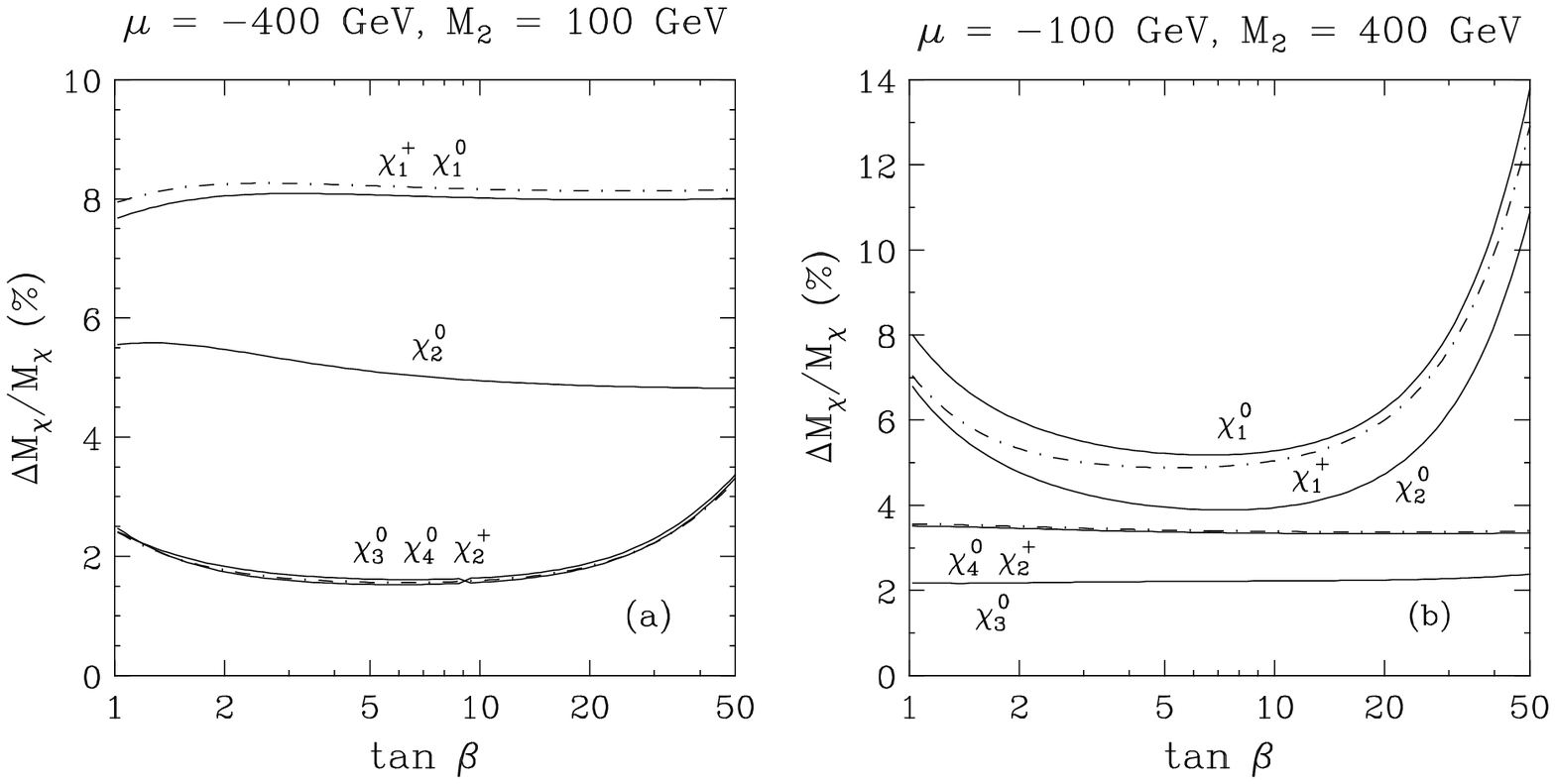}
\begin{center}
\parbox{5.5in}{
\caption[]{\small The relative difference (in percent)
between the tree level and one-loop level values of the chargino and
neutralino masses vs. $\tan\beta$ for $m_t=160$~GeV and $m_{sq}=1$~TeV.
(a) $\mu=-400$~GeV, $M_2=100$~GeV. (b) $\mu=-100$~GeV, $M_2=400$~GeV.
The chargino curves are indicated with dot-dashed lines.}}
\end{center}
\end{figure}
We find that the corrections can depend strongly on $\tan\beta$.
In Figs.6(a,b) we plot the relative correction
to the neutralino and chargino masses vs. $\tan\beta$
for two different choices of $M_2$ and $\mu$. We note that for a given
choice of the parameters $M_2$ and $\mu$
typically the lightest chargino and the two lightest neutralinos
will be predominantly Higgsinos, with the heavier particles being
predominantly gauginos, or vice versa. The relative correction
is fairly independent of $\tan\beta$ for the gauginos. However, when the
particle in question is mainly a Higgsino,
as we increase $\tan\beta$ from 1 to 2 the correction decreases as
the top Yukawa coupling decreases. For
$2\ \roughly{<}\ \tan\beta\ \roughly{<}\ 10$ the Higgsino corrections are
fairly constant. Increasing $\tan\beta$ further,
the bottom Yukawa becomes important
and $\Delta M_\chi/M_\chi$ increases again. The relative corrections
at $\tan\beta$=1 and $\tan\beta$=30 are about equal.

The gauge/Higgs/gaugino/Higgsino contributions to the corrections
are comparable to those of the matter sector in certain regions of
parameter space. For example, for $\mu\simeq -400$~GeV, $M_2\simeq 400$
GeV, $\tan\beta=1.5$ and $m_{sq}=1$ TeV, the correction due to the
matter sector is 5~GeV for $\chi_4^0$
and $\chi_2^+$, while the correction from the gauge/Higgs sector is about
7~GeV. Similarly, for $\chi_2^0$ and $\chi_1^+$ in this region the
matter correction is 7~GeV, while the gauge correction is 5 to 6~GeV.
The gauge/Higgs relative corrections are
small; in the experimentally allowed part of the
region $|\mu|<500$~GeV, $M_2<400$~GeV, we find
the corrections are usually in the range $-1$ to 1\%, and they
are always less than 3\%.

To conclude this section we list in Table 1 the lower bounds on the
neutralino and chargino masses, independent of $M_2,\ \mu,$ and
\begin{table}[htb]\centering
\begin{tabular}{|c||c|c|c|c|c|c|}\hline
  &  $\chi_1^0$ & $\chi_2^0$ & $\chi_3^0$
  &  $\chi_4^0$ & $\chi_1^+$ & $\chi_2^+$ \\ \hline
\hline $M_{\chi}^{\rm min}$ (GeV) & 16 & 44 & 70 & 108 & 47 & 99 \\
\hline\hline
${{\Delta M_\chi^{\rm min}\over
M_\chi^{\rm min}}\biggr|}_{m_t=160,m_{sq}=1  }$ (\%)
& $-18$ & $-1.2$ & 0.3 & 0.8 & 0 & 1.0 \\ \hline
${{\Delta M_\chi^{\rm min}\over
M_\chi^{\rm min}}\biggr|}_{m_t=200,m_{sq}=1.6}$ (\%)
& $-20$ & $-3.5$ & $-0.1$ & $-1.0$ & 0 & $-0.8$ \\ \hline
\end{tabular}
\parbox{5.5in}{
\caption[]{\small Lower bounds on the neutralino and chargino masses.
Row 1 lists the one-loop lower bounds in GeV with $m_t=160$~GeV and
$m_{sq}=1$~TeV. Row 2 shows
the percent difference between the lower bounds as determined at tree
and one-loop level. Row 3 is as row 2, with $m_t=200$~GeV
and $m_{sq}=1.6$~TeV.}}
\end{table}
$\tan\beta$ (we require $\tan\beta\geq1$). In the first row, we show the
lower bounds for the masses of the various particles at one-loop level,
with $m_t=160$~GeV and $m_{sq}$=1 TeV. We determine these
limits by finding the minimum $\chi$
mass in the experimentally allowed region of parameter space, where
both the masses and the boundary of the excluded region include the
one-loop mass corrections. In particular we have the limits from the
CDF cascade analysis $m_{\tilde{g}}>125$~GeV \cite{CDF,Hidaka,Haber},
and from LEP searches $M_{\chi_1^+}>47$~GeV \cite{LEP}.
We note that, excepting the lightest neutralino, these are essentially
the same limits found at tree
level in Ref.\cite{Hidaka}. In the second row of Table 1 we list the
relative difference (in percent) between our tree level
determination of the lower bounds and the one-loop results.
In particular, we find
that the minimum $\chi_1^0$ mass at tree level is 19.6~GeV.
Excepting $\chi_1^0$, the difference between the tree and one-loop
determinations are all about 1\% or less. These small differences
are the result of cancellations. Both the boundary of the excluded
region and the mass contours of all the particles shift in the same
direction, so that the net effect is small. This cancellation is
illustrated in Fig.4(a) of Ref.\cite{ourpaper}.
By way of contrast, for $\chi_2^0$ ($\chi_3^0$), the masses at tree
and one-loop level differ by 10\%\ (5\%) for fixed $\mu$, $M_2$
and $\tan\beta$ in the neighborhood of the minimum.
Because the contours all shift
relative to each other the cancellation is independent of the
parameters which affect the overall size of the corrections.
This is illustrated in the third row of Table 1, where we show the relative
difference between the tree level and one-loop level lower bounds,
setting $m_t=200$~GeV and $m_{sq}=1.6$~TeV.
Again, excepting $\chi_1^0$, the differences are small.

The cancellation of electroweak corrections is not apparent
for the lower bound deduced for the mass of the lightest neutralino
because it finds its minimum value
at a point in parameter space at which the gluino mass takes its lowest
value. Thus, because the gluino mass bound receives
strong corrections as discussed in Section 3 it is clear that the
cancellations cannot occur for $\chi_1^0$. However, there is some
cancellation taking place. For a gluino mass correction of 30\%,
the corresponding 30\% shift in $M_2$ (and $M_1$) changes the tree
level $\chi_1^0$ mass by about 27\%. The one-loop correction to
the $\chi_1^0$ mass partially compensates for this correction,
and the net effect is about a 20\% shift.

The chargino and neutralino mass limits are not likely to increase much
in the near future. Excepting the lightest neutralino, all of the
particles find their minimum values at points in parameter space
essentially at the kinematic limit of LEP. Additionally, the points are
far from the gluino mass bound. The $\chi_1^0$ mass limit increases
proportionally to the gluino mass bound as $m_{\chi_1^0}^{\rm min}
\simeq m_{\tilde{g}}^{\rm min}/7.8$ for gluino mass limits up to
155~GeV. The $\chi_2^0$ mass limit remains unaffected for gluino mass
bounds up to 180~GeV, the $\chi_4^0$ and $\chi_2^+$ limits are
unaffected for gluino mass bounds up to 220~GeV, and the
$\chi_3^0$ limit is independent of the gluino mass limit up to
$m_{\tilde{g}}^{\rm min}=400$~GeV.

\section{Conclusions}

We have computed all of the one-loop radiative corrections to the gaugino
and Higgsino masses in the MSSM. We performed the
diagrammatic calculation in the
on-mass-shell renormalization scheme.

We first considered the
radiatively generated mass for a gluino which is massless at tree level.
This mass falls generally in the ``light gluino window'',
$m_{\tilde{g}}\ \roughly{<}\ $3~GeV. There is by no means a consensus in the
literature as to whether gluinos in this mass range have already been
excluded by current experimental data. If such a light gluino
were discovered,
it would be compelling evidence for the condition $m_{1/2}=0$ at the GUT
scale, and it would greatly constrain the parameters in the top/stop
sector of the MSSM.

Next, we discussed the corrections to the mass of a
massive gluino. We found large $\cal O$(20\%) corrections
for values of the gluino mass below the quark/squark thresholds, and
small ${\cal O}(1\%)$ corrections for gluino masses above the
quark/squark thresholds. These corrections
must be considered when relating the gluino mass to the mass parameters
$M_1$ and $M_2$ which enter into the evaluation of the chargino and
neutralino masses.

Lastly, we discussed the results for the corrections to the
neutralino and chargino masses. While the corrections from the
gauge/Higgs sector can be as large as those from the matter sector,
the gauge/Higgs sector corrections are never as large as 3\%. The complete
corrections are typically 3-8\%. These corrections are relevant
when determining the values of the underlying parameters from a
set of physical $\chi$ masses. The underlying parameters can then be
used to constrain or make predictions for other sectors of the theory.
For example, $\mu$ enters into the Higgs boson and heavy quark/squark
sectors, and $\tan\beta$ enters into the Higgs, fermion/sfermion and
quark/squark sectors. Alternatively, given a set of underlying parameters
$\mu,\ m_{1/2},$ etc., at some high (e.g. GUT) scale, the parameters
can be
determined at a low scale via the renormalization group equations and
the corrections presented here can then determine the physical spectrum
of gaugino and Higgsino masses. Of course the physical spectrum is
independent of the scale at which the RGE evolution is terminated,
up to higher orders in perturbation theory.

We showed how the presently excluded region of parameter space
is affected when the mass corrections are considered, and we found the
one-loop lower bounds on the neutralino and chargino masses.
Excepting the lightest neutralino,
the difference between the tree and one-loop determinations of the lower
bounds were found to be small ($\sim1\%$), due to cancellations.
The $\chi_1^0$ lower bound receives 20\% corrections.

\vskip 1cm
\noindent {\Large\bf Appendix}
\vskip .7cm
\setcounter{equation}{0}
\renewcommand{\theequation}{A\arabic{equation}}

Below we list the relevant one loop
self-energies for the charginos and neutralinos, including only
the gauge/Higgs particles and their partners in loop. The  matter loop
contributions can be found in Ref. \cite{ourpaper}.
The functions $B_0,\ B_1$ are given in Ref.\cite{Passarino}, except that
here we adopt the Minkowski metric $(+,-,-,-)$.
The various $A$ and $B$ couplings in the following expressions are
the vector and axial vector (scalar and axial scalar)
couplings of the charginos and neutralinos to the gauge bosons
(Higgs particles). They can be found in Refs.\cite{Haber&Kane,HiggsGuide}.
The indices $i,j$ refer to charginos while the indices $\alpha, \beta$
refer to neutralinos.
The $H_n^0,\ n=1..4$ denote the neutral Higgs bosons,
$H,h,A,G^0$, while $H_m^+,\ m=1,2,$ denote the charged Higgs bosons,
$H^+,$ and $G^+$. The calculation was performed in the Feynman gauge,
in which $M_{G^0}=M_Z$ and $M_{G^+}=M_W$. Dimensional reduction\cite{Siegel}
was used in regularizing the theory.

We have for the charginos
\begin{eqnarray}
(4\pi)^2\ \Sigma^+_{1_{ii}}(p^2)  = &&
            \sum_{j=1}^{2}\sum_{n=1}^{4}M_{\chi^+_j}
            \left((A_{ijn})^2-(B_{ijn})^2\right)
            B_0(p^2, M_{\chi^+_j}^2, M_{H_n^0}^2)\nonumber\\
            & + & \sum_{\beta=1}^{4}\sum_{m=1}^{2} M_{\chi^0_{\beta}}
            \left(|A_{\beta im}|^2-|B_{\beta im}|^2\right)
            B_0(p^2, M_{\chi^0_{\beta}}^2, M_{H_m^+}^2) \nonumber\\
            & - &4 \sum_{j=1}^{2}M_{\chi^+_j}
            \left((A_{ijZ})^2- (B_{ijZ})^2\right)
            B_0(p^2, M_{\chi^+_j}^2, M_Z^2) \\
            & - &4 e^2 M_{\chi^+_i}B_0(p^2, M_{\chi^+_i}^2,0)\nonumber\\
            & - &4\sum_{\beta=1}^{4}M_{\chi^0_{\beta}}
            \left(|A_{\beta i Z}|^2- |B_{\beta i Z}|^2\right)
      B_0(p^2, M_{\chi^0_{\beta}}^2, M_W^2) \nonumber
\end{eqnarray}
\pageb
\begin{eqnarray}
(4\pi)^2\ \Sigma^+_{\gamma_{ii}}(p^2)  = &&
             \sum_{j=1}^{2}\sum_{n=1}^{4}
            \left((A_{ijn})^2+(B_{ijn})^2\right)
            B_1(p^2, M_{\chi^+_j}^2, M_{H_n^0}^2)\nonumber\\
            & + & \sum_{\beta=1}^{4}\sum_{m=1}^{2}
            \left(|A_{\beta im}|^2+|B_{\beta im}|^2\right)
            B_1(p^2, M_{\chi^0_{\beta}}^2, M_{H_m^+}^2) \nonumber\\
            & + &2 \sum_{j=1}^{2}
            \left((A_{ijZ})^2+ (B_{ijZ})^2\right)
            B_1(p^2, M_{\chi^+_j}^2, M_Z^2) \\
      & + &2 e^2 B_1(p^2, M_{\chi^+_i}^2, 0)\nonumber\\
            & + &2\sum_{\beta=1}^4
            \left(|A_{\beta i W}|^2+ |B_{\beta i W}|^2\right)
      B_1(p^2, M_{\chi^0_{\beta}}^2, M_W^2) \nonumber
\end{eqnarray}

\noindent The corresponding expressions for the neutralinos are
\begin{eqnarray}
(4\pi)^2\ \Sigma^0_{1_{\alpha\alpha}}(p^2)  = &&
            \sum_{\beta=1}^{4}\sum_{n=1}^{4}M_{\chi^0_{\beta}}
            \left(|A_{\alpha\beta n}|^2-|B_{\alpha\beta n}|^2\right)
            B_0(p^2, M_{\chi^0_{\beta}}^2, M_{H_n^0}^2)\nonumber\\
            & + & 2\sum_{j=1}^{2}\sum_{m=1}^{2} M_{\chi^+_j}
            \left(|A_{\alpha jm}|^2-|B_{\alpha jm}|^2\right)
            B_0(p^2, M_{\chi^+_j}^2, M_{H_m^+}^2) \\
            & - &4 \sum_{\beta=1}^4M_{\chi^0_{\beta}}
            \left(|A_{\alpha\beta Z}|^2- |B_{\alpha\beta Z}|^2\right)
            B_0(p^2, M_{\chi^0_{\beta}}^2, M_Z^2) \nonumber\\
            & - &8\sum_{j=1}^{2}M_{\chi^+_j}
            \left(|A_{\alpha j W}|^2-|B_{\alpha j W}|^2\right)
      B_0(p^2, M_{\chi^+_j}^2, M_W^2) \nonumber
\end{eqnarray}

\begin{eqnarray}
(4\pi)^2\ \Sigma^0_{\gamma_{\alpha\alpha}}(p^2)  = &&
         \sum_{\beta=4}^{2}\sum_{n=1}^{4}
         \left(|A_{\alpha\beta n}|^2+|B_{\alpha\beta n}|^2\right)
         B_1(p^2, M_{\chi^0_{\beta}}^2, M_{H_n^0}^2)\nonumber\\
         & + & 2\sum_{j=2}^{4}\sum_{m=1}^{2}
         \left(|A_{\alpha  jm}|^2+|B_{\alpha jm}|^2\right)
         B_1(p^2, M_{\chi^+_j}^2, M_{H_m^+}^2) \\
         & + &2 \sum_{\beta=1}^{4}
            \left(|A_{\alpha\beta Z}|^2+|B_{\alpha\beta Z}|^2\right)
            B_1(p^2, M_{\chi^0_{\beta}}^2, M_Z^2) \nonumber\\
            & + &4\sum_{j=1}^{2}
            \left(|A_{\alpha jW}|^2+|B_{\alpha jW}|^2\right)
      B_1(p^2, M_{\chi^+_j}^2, M_W^2) \nonumber
\end{eqnarray}

\vskip1cm
\noindent {\Large\bf Acknowledgments}
\vskip .5cm
\noindent DP acknowledges the support of the National Science Foundation under
grant NSF-PHY-90-9619 and the state of Texas under grant TNRLC-RGFY-93-292.
AP acknowledges the support of the U.S. Department of Energy under contract
DE-ACO2-76ER01428 (Task B).

\pageb

\end{document}